\documentstyle[preprint,prc,aps,epsfig]{revtex}

\begin{document}
\title{An Analysis of the $^{12}$C+$^{12}$C Reaction
Using a New Type of Coupling Potential}
\author{I. Boztosun \footnote[1]{Present address: Computational Mathematics Group,
University of Portsmouth, Mercantile House, Portsmouth PO1 2EG UK}
\footnote[2]{Permanent address :
Department of Physics, Erciyes University, Kayseri 38039 Turkey} and W.D.M. Rae}
\address{Department of Nuclear Physics, University of Oxford,
Keble Road, Oxford OX1 3RH UK}

\date{\today}
\maketitle

\begin{abstract}
A new approach has been used to explain the experimental data for the
$^{12}$C+$^{12}$C system over a wide energy range in the laboratory
system from 32.0 MeV to 126.7 MeV. This new
coupled-channels based approach involves replacing the usual first
derivative coupling potential by a new, second-derivative coupling
potential. This paper first shows and discusses
the limitation of the standard coupled-channels theory in the case
where one of the nuclei in the reaction is strongly deformed. Then,
this new approach is shown to improve consistently the agreement with
the experimental data: the elastic scattering, single-2$^{+}$ and
mutual-2$^{+}$ excitation inelastic scattering data
as well as their 90$^{\circ}$
elastic and inelastic excitation functions with little energy-dependent
potentials. This new approach makes
major improvement on all the previous coupled-channels calculations
for this system.
\end{abstract}



\section
{Introduction}
Forty years ago, it was observed that the elastic cross-section of
the $^{12}$C+$^{12}$C system varies rapidly
with bombarding energy. This structure in the excitation functions,
which could also be observed in other systems such as
$^{12}$C+$^{16}$O and $^{16}$O+$^{16}$O, has remained a subject
attracting continuous interest from both theoretical and experimental
points of views. Consequently, a large body of data over a wide
energy range has been accumulated for the $^{12}$C+$^{12}$C system
from the systematic studies of this reaction \cite{Bro60,Sto77,Sto79,Led83}.

However, there has been no global model that describes consistently
the available elastic and inelastic scattering data over a wide energy
range and this reaction presents a challenge to the many different
theoretical models. Some of the problems can be summarised as:
$(1)$ no consistent description of the elastic scattering, single-2$^{+}$
and mutual-2$^{+}$ excitation inelastic scattering data as well as their
90$^{\circ}$ excitation function;
$(2)$ the out of phase problem between the theoretical predictions
and the experimental data for these states;
$(3)$ no simultaneous description of the individual angular
distributions and resonances;
$(4)$ the magnitude of the mutual-2$^{+}$ excitation inelastic scattering
data is unaccounted for.

The elastic scattering data of this system has been studied
systematically and progress has been made using the optical model
(see the review by Brandan and Satchler \cite{Bra97}). However,
the inelastic scattering has received little attention and there
is no systematic study over a wide energy range and the
above-mentioned problems could not be explained using the standard
coupled-channels models (see for example \cite{Sto79,Wol82,Fry97,Rae97,Sak99,Ito99,Kub85}).

Stokstad {\it et al} \cite{Sto79} were the first to study the elastic
and single-2$^{+}$ excitation inelastic scattering data using the DWBA
and the coupled-channels methods
from $E_{Lab}$=74.2 MeV to 126.7 MeV. They obtained reasonable
agreement with the elastic scattering data.
However, they could not reproduce the correct oscillatory structure
for the single-2$^{+}$ excitation inelastic scattering data and the magnitude
of the data could not be accounted for correctly. They did not
study the mutual-2$^{+}$ excitation inelastic scattering data in their calculations.
No theoretical calculations has predicted the magnitude of this
data correctly over a wide energy range.

Wolf, Satchler and others \cite{Wol82} studied this
system at three different energies. They used a double-folding
potential and an angular-momentum dependent imaginary potential
in their coupled-channels calculations. They could not reproduce the
experimental data measured at $E_{Lab}$=74.2, 93.8, and 126.7 MeV.
In particular, the theoretical predictions for the mutual-2$^{+}$
excitation inelastic scattering data were very small by
factors of 3 to 10 with respect to the experimental data. The
results of the single-2$^{+}$ excitation inelastic
scattering calculations were also very oscillatory in
comparison with the experimental one.
We encountered the same problems in our standard coupled-channels
calculations. Varying the parameters and changing the shape of the
real and imaginary potentials do not provide a solution, as
discussed in section \ref{res}.

Fry {\it et al} \cite{Fry97,Rae97} also worked on this reaction to
obtain the integrated cross-section (also known as Cormier's
resonances) for the single-2$^{+}$ and mutual-2$^{+}$ excitation
channels using the coupled-channels method. They made use of a double-folding
potential like the one of Stokstad {\it et al} \cite{Sto79} and an
angular momentum-dependent imaginary potential. However, this method
totally failed and no improvement of the densities in the
double-folding potential would solve the magnitude problem of the
mutual-$2^{+}$ excitation inelastic scattering data.
The same problems are observed in other
authors' works such as  Sakuragi \cite{Sak99} and Ito \cite{Ito99}.

Another interesting analysis was made by Ordo\~{n}ez {\it et al}
\cite{Ord 86}. They showed the necessity of using a real potential that
has a minimum in the surface region. They reported a detailed
phase-shift analysis of the $^{12}$C+$^{12}$C elastic scattering data
in the range of 11.0$\le$$E_{Lab}$$\le$66.0 MeV. This analysis
revealed a striking sequence of  gross structure resonances that
appeared to form a rotational band from $l$=0 to at least 16$\hbar$.
These resonances were simulated by shape-resonances in a real
potential with a secondary minimum at large radii related to the
shape-isomeric doorway states in $^{24}$Mg.

The interesting feature of their work is the double-peaked nature of the
real potential. It is clear that this potential
does explain the resonance data, which
other models have failed to reproduce within such a large energy range.
As will be argued in section \ref{newcc}, there is a resemblance between this
potential and our total nuclear potential ($V_{real}$+$V_{coupling}$).

Ordo\~{n}ez {\it et al} could not justify this double-peaked
potential, other than by asserting it was required to fit the
experimental data. This paper and a forthcoming paper \cite{Boz3} shall argue
that this deepening at
the surface is due to the strongly deformed structure of the $^{12}$C
and may indicate a super-deformed state of the compound nucleus,
$^{24}$Mg. It is also clear that Ordo\~{n}ez
{\it et al} took into account the coupling effects in their
optical model calculations by introducing such a deepening at
the surface without running coupled-channels calculations.

The literature clearly shows that the standard coupled-channels
approach can fit neither any of the individual angular
distributions nor the 90$^{\circ}$ elastic scattering excitation
function {\it simultaneously}. For the resonance calculations, the
situation is the same. That is, even if one fitted the Cormier's
resonances observed for the single-2$^{+}$ and mutual-2$^{+}$ excitation channels,
it would be, at the same time, impossible to fit the 90$^{\circ}$
elastic scattering excitation function. Clearly, the
$^{12}$C+$^{12}$C system has numerous problems to which no consistent
global solution has been provided yet.

The overview of previous works indicates that the central potentials
are actually quite reasonable since they have given the resonances at
the correct energies and with sensible widths.  Within the optical model
calculations, they have also given
very good agreements for the elastic scattering angular distributions
or the 90$^{\circ}$ elastic scattering excitation functions
independently. However, the calculations for the mutual-2$^{+}$ excitation
inelastic scattering data is
in general under-predicted by a large factor and the oscillatory
structure of the data can not be reproduced correctly.
They have remained unsolved so far.

Therefore, our aim of analysing the $^{12}$C+$^{12}$C system is to
search for a global solution for some of these problems with little
energy-dependent potentials within the coupled-channels formalism
from 32.0 MeV to 126.7 MeV in the laboratory system.

In the next section, we introduce the model potentials used to analyse
the experimental data and the results of these analyses are shown in
section \ref{res}, where we also make a discussion of the limitations
of the standard coupled-channels method and higlight the problems.
Section \ref{newcc} is devoted to the analyses of the experimental
data using our new coupling potential and the results are shown in
section \ref{resnew}. Finally, the section \ref{summary} gives a
summary and a discussion of the new and standard coupled-channels calculation.

\section{The Standard Coupled-Channels Calculations}
\label{stan}
A recent critical review  by Kond\=o {\it et al} \cite{Kon 98} found
that a potential with a real depth of $\sim$300 MeV was able to
account for the 90$^{\circ}$ elastic excitation function at
low energies ($E_{Lab}$$\leq$75.0 MeV). The real potentials proposed in
this paper are tested and their parameters have to be readjusted due to
the coupling effects in the coupled-channels calculations.

In our coupled-channels calculations,
the interaction between the $^{12}$C nuclei is described by
a deformed optical potential. The real potential has
the square of a Woods-Saxon shape:
\begin{equation}
V_{N}(r) = \frac{-V_{N}}{(1+exp(r-R_{N})/a_{N})^{2}}
\label{realpot}
\end{equation}
and the parameters, as shown in table \ref{param}, are fixed to reproduce
the 90$^{\circ}$ elastic scattering excitation function. The Coulomb potential
is assumed to be that of a uniformly charged nucleus with a radius of 5.5 fm.

The imaginary potential has the standard Woods-Saxon volume shape:
\begin{equation}
W(r)=-\frac{W}{1+exp((r-R_{W})/a_{W})}
\label{imagpot}
\end{equation}
and its depth increases quadratically with energy as:
\begin{equation}
W = -2.69+0.145E_{Lab}+0.00091E_{Lab}^{2}
\label{imag}
\end{equation}
The parameters of the radius and diffuseness are shown in table \ref{param}.

Since the $^{12}$C nucleus is strongly deformed, its collective excitation
has been treated in the framework of the coupled-channels formalism.
The $^{12}$C nucleus has a static quadrupole deformation, which is
taken into account by deforming the real optical potential with a Taylor
expansion about R=$R_{0}$ in the usual way \cite{Sat83}:
\begin{equation}
U(r-R)=U(r-R_{0})-\delta R \frac{\partial}{\partial R_{0}}U(r-R_{0})+
\frac{1}{2!}(\delta R)^{2}
\frac{\partial^{2}}{\partial R_{0}^{2}}U(r-R_{0})- \ldots
\end{equation}
For the projectile $P$ and the target $T$

\begin{eqnarray}
\delta R_{P}=R_{P}\beta_{2}Y_{20}(\theta,\phi)
\nonumber \\
\delta R_{T}=R_{T}\beta_{2}Y_{20}(\theta,\phi).
\label{pot}
\end{eqnarray}
$R_{P}$ and $R_{T}$ are the radii of the projectile and target.
The form factors \cite{Sat83} are

\begin{equation}
F_{P}(r)=R_{P} \left[\frac{\partial}{\partial R_{0}}U(r,R_{0}) \right], \hspace{2cm}
F_{T}(r)=R_{T} \left[\frac{\partial}{\partial R_{0}}U(r,R_{0}) \right]
\label{f1}
\end{equation}
\begin{equation}
H_{P,T}(r)=
\frac{1}{(4\pi)^{1/2}}R_{P}R_{T} \frac{\partial^{2}U(r,R_{0})}
{\partial R_{0}^{2}}
\label{f2}
\end{equation}

$F_{P}(r)$ and $F_{T}(r)$ in equation \ref{f1} are the first-order form factors that
account for the excitations of the projectile and
target nuclei, while $H_{P,T}(r)$ in equation \ref{f2} is the second order form
factor that accounts for their mutual excitation.

In equation (\ref{pot}), $\beta_{2}$=-0.6 is the deformation parameter of
the $^{12}$C nucleus. This empirical value is derived from its known B(E2) value.
The value of B(E2) is 42 e$^{2}fm^{4}$ \cite{Ste 66}.
(A more recent measurement gives an average value of 39$\pm$4
e$^{2}fm^{4}$ \cite{Ajz 75}).

\begin{figure}
\epsfxsize=10.0cm
\vskip-2.5cm
\centerline{\epsfbox{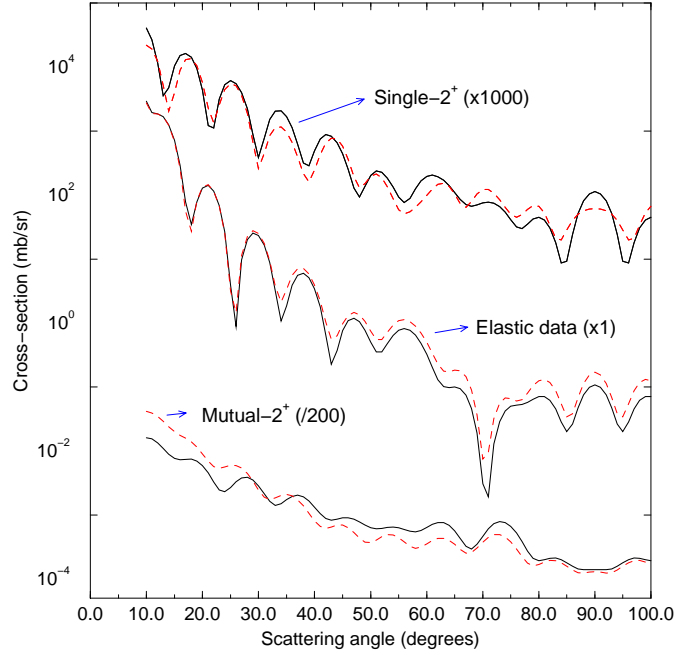}}
\vskip-1.0cm
    \caption{A comparison of the results of the simultaneous
mutual excitation (the dashed line) and
the sequential one (the solid line) for the elastic,
single-2$^{+}$ and mutual-2$^{+}$ excitations at $E_{Lab}$=93.8 MeV.}
\label{mutualeffect}
\end{figure}

In the standard coupled-channels calculations of inelastic scattering involving mutual
excitation  of the two nuclei, the codes CHUCK \cite{Kunz} and FRESCO
\cite{Tho88} are used in such a way that the two nuclei are excited sequentially.
However, we think it essential that {\it simultaneous} mutual excitation of the two
nuclei be included in the calculations. To do so, we use the mutual-2$^{+}$ excitation
inelastic scattering data  that are available. We modify the code CHUCK to take into
account the {\it simultaneous} mutual excitation process \cite{Boz1}. It is observed
that the simultaneous mutual excitation of the two nuclei does affect the calculations,
in particular in the resonance region where the calculations are very sensitive to the
small variations of the potential parameters. This is demonstrated in figure \ref{mutualeffect}
at $E_{Lab}$=93.8 MeV since we have available experimental data for all the states
considered in this paper.
\section{Results}
\label{res}

\begin{figure}
\epsfxsize=12.0cm
\vskip-6.5cm
\centerline{\epsfbox{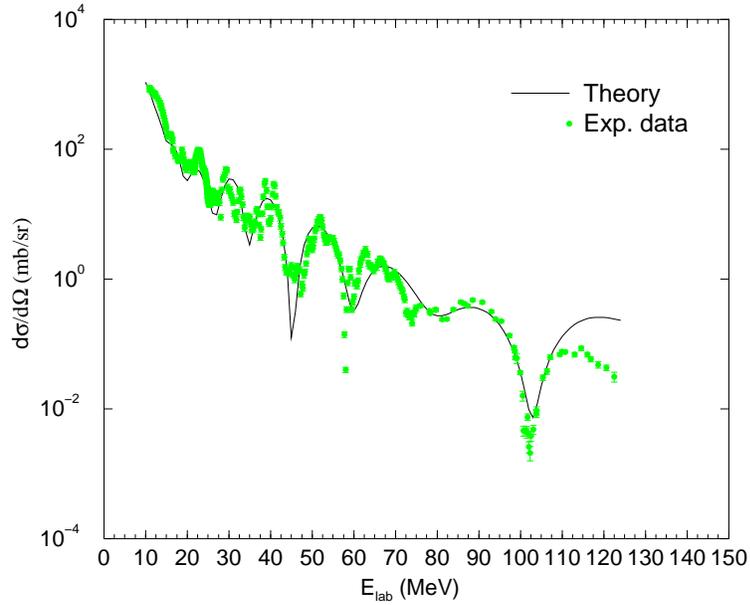}}
\vskip-1.5cm
    \caption{The comparison of the experimental data and the
results of the standard coupled-channels calculation for the 90$^{\circ}$ elastic scattering excitation function.}
\label{exc1}
\end{figure}

The result of the 90$^{\circ}$ elastic scattering excitation function obtained
using the parameters of table \ref{param} is shown in figure \ref{exc1}.
The theoretical predictions and the experimental data are in very
good agreement, but, as Kondo {\it et al} found, this potential family does not
fit the individual elastic scattering and inelastic scattering data as well as
their excitation functions {\it simultaneously}.

We have attempted to obtain reasonable fits to the individual
angular distributions by changing the parameters of the real potential,
shown in table \ref{param}, but without success.
Some authors \cite{Sak99,Bra90} also found a potential family that reproduces
the individual angular distributions, but does not fit the
90$^{\circ}$ elastic scattering excitation function.

\begin{figure}[bt]
\epsfxsize=14.5cm \centerline{\epsfbox{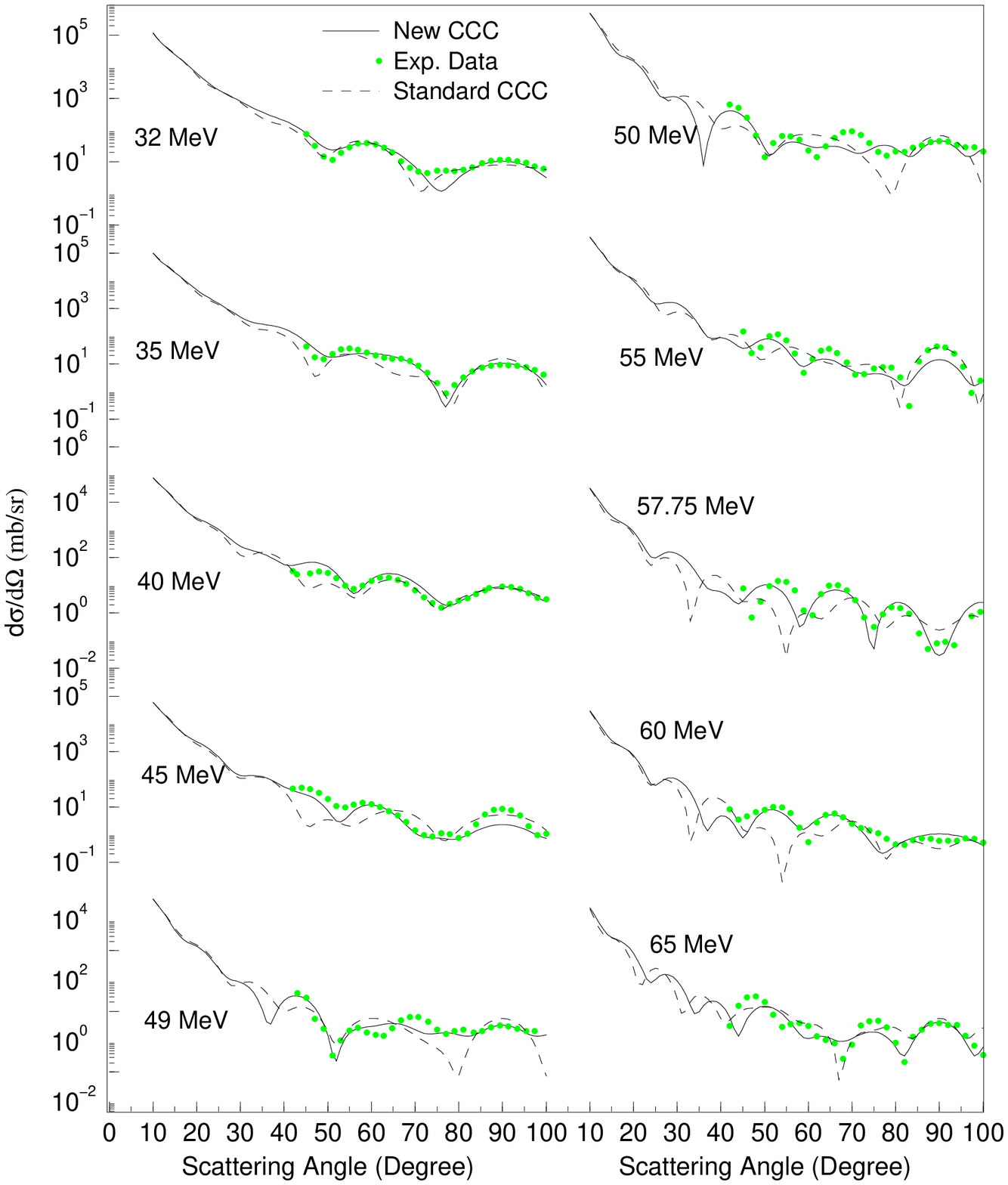}}
    \caption{Ground state results: The dashed lines show the predictions of the standard coupled-channels calculations (see table \ref{paramws2} for the parameters) while the solid lines show the results of the new coupled-channels calculations, obtained using new coupling potential with the empirical $\beta$ value ($\beta_{2}^{C}$=$\beta_{2}^{N}$=-0.6)  (see table \ref{param2nd} for the parameters).}
\label{ground1}
\end{figure}

\begin{figure}[bt]
\epsfxsize 14.5cm \centerline{\epsfbox{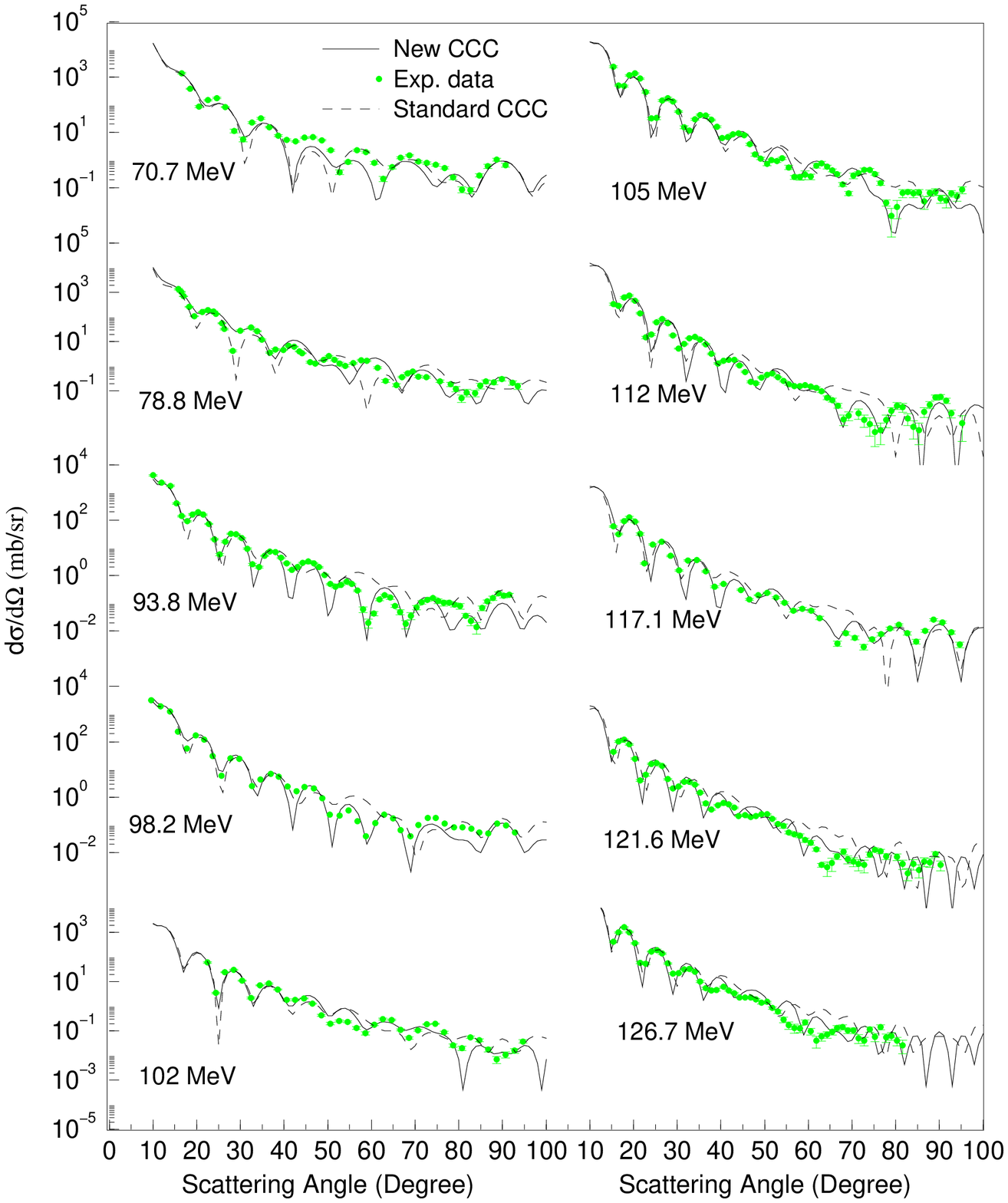}}
    \caption{Ground state results: The dashed lines show the predictions of the standard coupled-channels calculations (see table \ref{paramws2} for the parameters) while the solid lines show the results of the new coupled-channels calculations, obtained using new coupling potential with the empirical $\beta$ value ($\beta_{2}^{C}$=$\beta_{2}^{N}$=-0.6)  (see table \ref{param2nd} for the parameters) ({\it continued from figure \ref{ground1}}).}
\label{ground2}
\end{figure}

\begin{figure}
\epsfxsize 15.5cm \centerline{\epsfbox{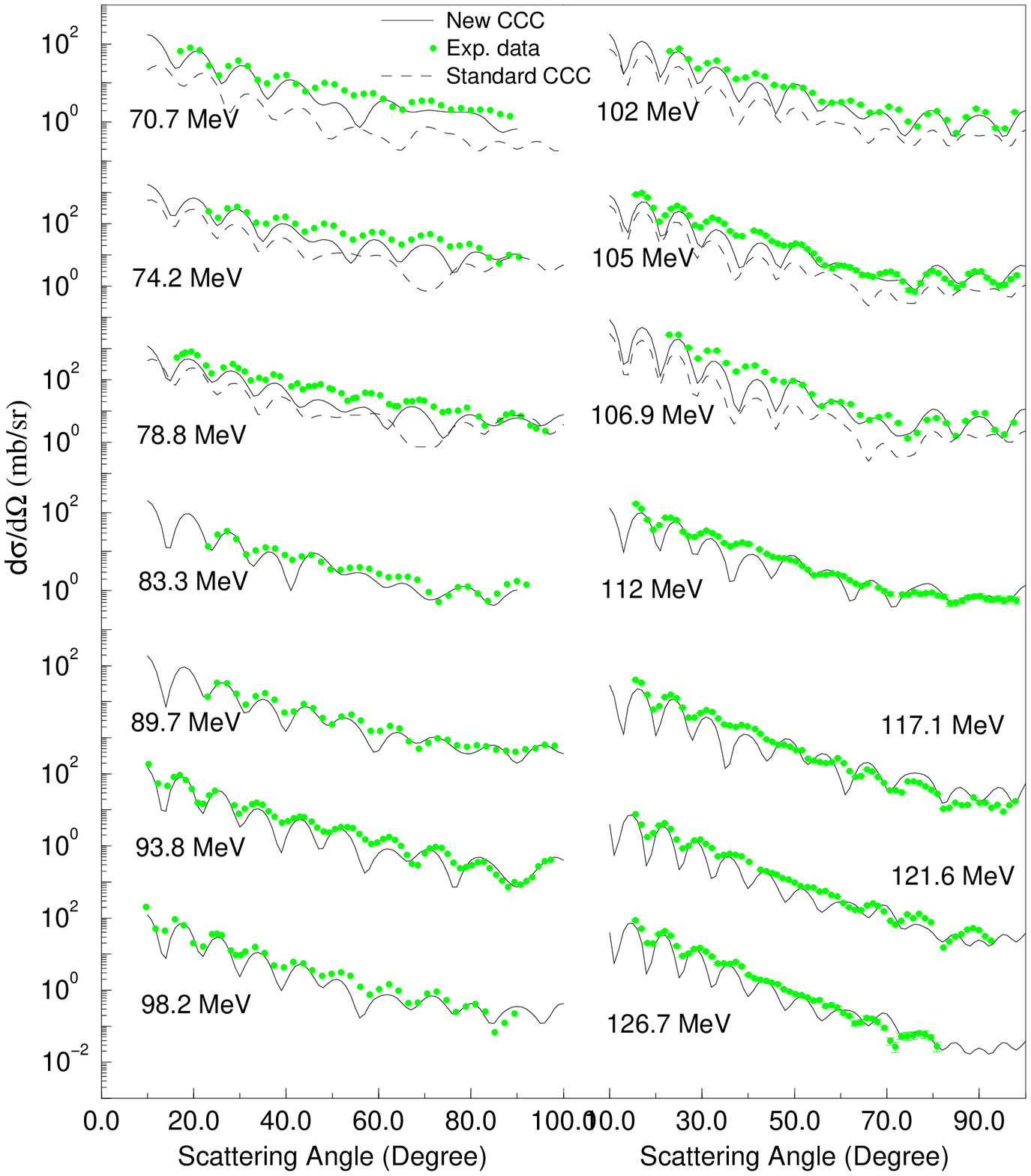}}
\vskip-1.0cm
    \caption{Single-2$^{+}$ state results: The dashed lines show the predictions of the standard coupled-channels calculations (see table \ref{paramws2} for the parameters) while the solid lines show the results of the new coupled-channels calculations, obtained using new coupling potential with the empirical $\beta$ value ($\beta_{2}^{C}$=$\beta_{2}^{N}$=-0.6)  (see table \ref{param2nd} for the parameters).}
\label{single}
\end{figure}

\begin{figure}
\vskip-3.0cm
\epsfxsize 15.5cm \centerline{\epsfbox{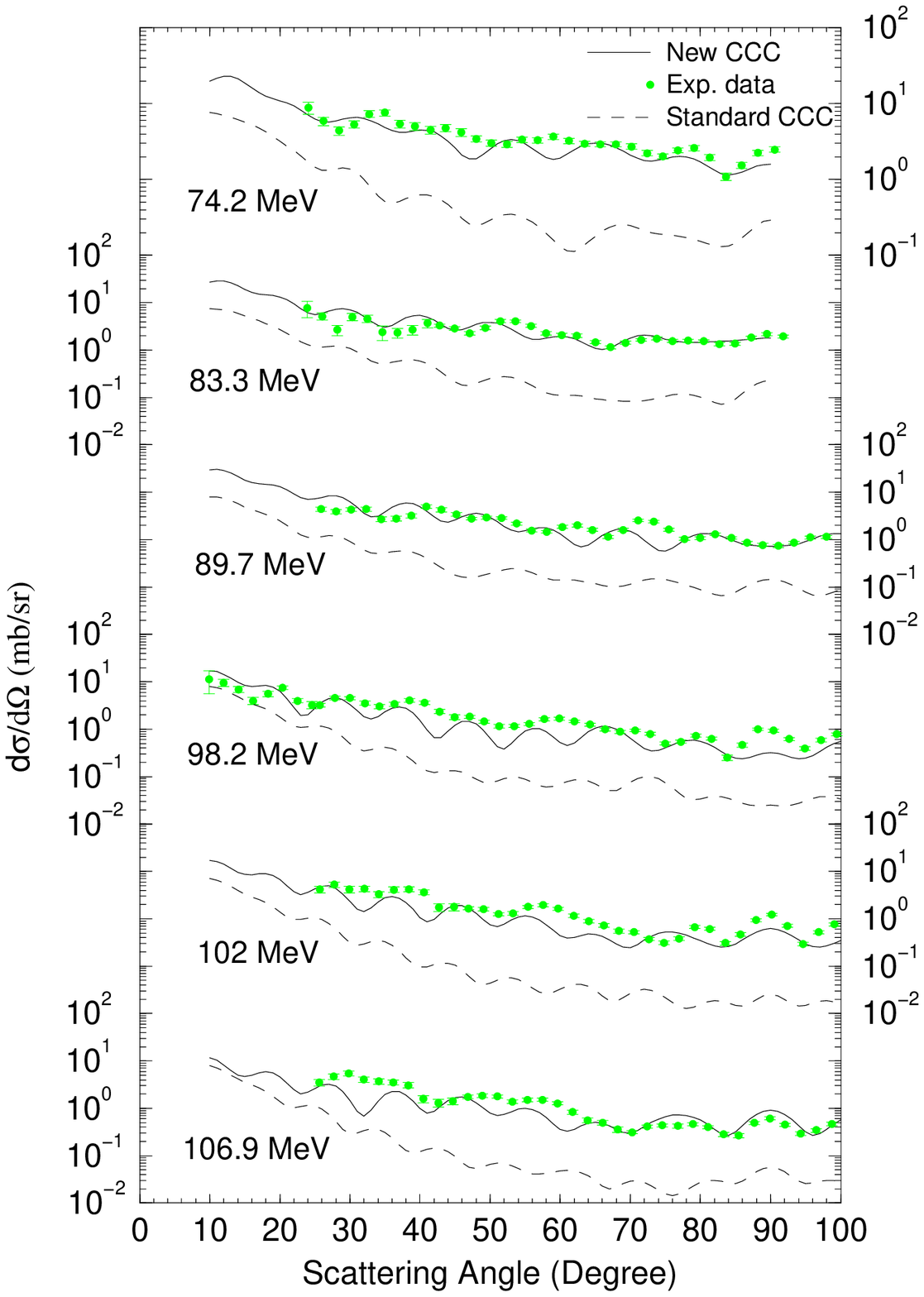}}
\vskip-0.0cm
    \caption{Mutual-2$^{+}$ state results: The dashed lines show the predictions of the standard coupled-channels calculations (see table \ref{paramws2} for the parameters) while the solid lines show the results of the new coupled-channels calculations, obtained using new coupling potential with the empirical $\beta$ value ($\beta_{2}^{C}$=$\beta_{2}^{N}$=-0.6)  (see table \ref{param2nd} for the parameters).}
\label{mutual}
\end{figure}

To overcome this difficulty, we searched for a new potential family
by readjusting the parameters of the real potential and letting the
imaginary potential change freely. The parameters are shown in
table \ref{paramws2}. Except in the resonance
regions, we obtained satisfactory agreement for the
elastic scattering data as shown in figures \ref{ground1} and \ref{ground2}
with dashed lines. However, the theoretical predictions of the
magnitudes and the phase of the oscillations are not in good
agreement with the experimental data for the single-2$^{+}$ state, as shown
in figure \ref{single} with dashed lines.
The out-of-phase and magnitude problems between the theoretical
calculations and the
experimental data are clearly seen at many energies. These results for
the elastic and single-2$^{+}$ excitation inelastic scattering are almost
identical to those obtained by Stokstad {\it et al} \cite{Sto79}. For the
mutual-2$^{+}$ excitation inelastic scattering data,  as shown in
figure \ref{mutual} with dashed lines, there is
no agreement and the theoretical predictions of the
magnitude of mutual-2$^{+}$ excitation inelastic scattering data are much
smaller than the experimental one; they are under-predicted
by a factor of 3 to 10. Nevertheless, our results for the
mutual-2$^{+}$ excitation inelastic scattering
data are in conformity with the findings of the references
\cite{Wol82,Fry97,Rae97,Sak99,Kub85}, a problem
mentioned by many authors in a recent international conference on clustering
(ICC '99) \cite{Sak99,Ito99,Abe99,Boz99}. In order to make a comparison
with the new calculations, presented in the next section, some of the results for the
single-2$^{+}$ and mutual-2$^{+}$ states are shown in figure
\ref{single} and \ref{mutual}.

We had anticipated that the inclusion of the simultaneous
mutual excitation of two nuclei could solve the magnitude
problem of the mutual-2$^{+}$ excitation data. However,
although this effect has improved the details of the fits
to the experimental data, it failed to provide a solution.
The magnitude of the mutual-2$^{+}$ excited state
cross-section is still one of the major outstanding problems
of this reaction.

In the past, the magnitude problem for the single-2$^{+}$ excitation
inelastic scattering calculations was solved  for different reactions by changing the
empirical $\beta$ value \cite{Abe78,Sci97}. Thus, the same solution
was expected to apply to the $^{12}$C+$^{12}$C
system for the single-2$^{+}$ and mutual-2$^{+}$ excitations inelastic
scattering calculations. For this purpose, we increased
the $\beta$ value to -1.2,
which is twice the actual value and has no physical justification.
However, although the agreement between theoretical predictions and the
experimental data for the magnitudes of the
single-2$^{+}$ and mutual-2$^{+}$ excitations inelastic scattering data is
improved, the theoretical predictions for the elastic scattering data are very
poor; the same holds for the 90$^{\circ}$ elastic scattering excitation
functions.

Within the coupled-channels formalism, the reason for this failure may be
understood if the effect of
changing the real potential on the inelastic scattering cross-section
is considered. The method of obtaining the coupling potential that
describes the inelastic scattering has been based on perturbation theory.
Since the coupling potential is connected to
the real term by a Taylor expansion around the surface of the nucleus, changing
the real potential has a substantial effect on the elastic
scattering data, but not on the inelastic scattering one.
Therefore, according to
this standard procedure, the coupling potential has the same
energy-dependence as the central term. Actually, Smithson {\it et al}
\cite{Smi 90} analysed the inelastic scattering data for the
$^{16}$O+$^{208}$Pb system and asserted that the standard
deformation procedure is inadequate for the description of the
inelastic scattering data. They  also concluded that there is no
reason for the coupling potential to have the same energy dependence as the central potential.
\section{New Coupling Potential}
\label{newcc}
If we consider two $^{12}$C nuclei approaching each other,
the double-folding model will generate an {\it oblate} potential
which is correct at large distance. When these two nuclei
come close enough, they create the compound nucleus $^{24}$Mg which
is a {\it  prolate} nucleus in its ground state, whereas the folding
model yields an oblate (attractive) potential in this case. How well
the double-folding model describes a prolate nucleus
with an oblate potential is unclear and this may be the reason why the earlier calculations using
a double-folding model in the coupled-channels method were unable to
provide a consistent solution to the problems of this reaction.

\begin{figure}
\epsfxsize=9.0cm
\centerline{\epsfbox{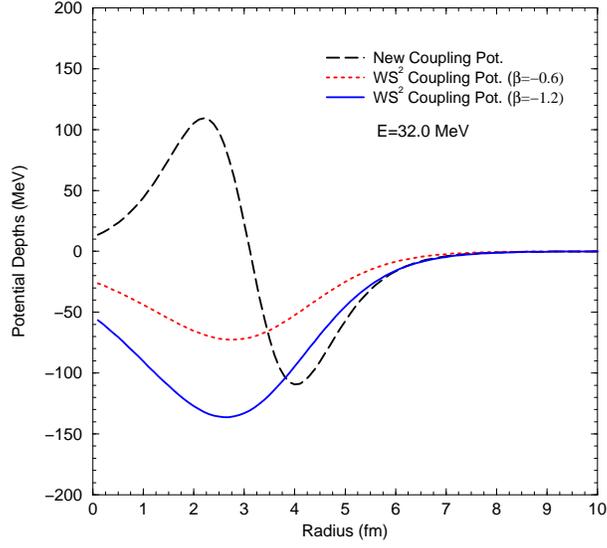}}
\vskip-3.5cm
\caption{The comparison of the {\it standard
coupling potential} $(1)$ with $\beta$=-0.6, $(2)$ with $\beta$=-1.2 and
our {\it new coupling potential} for $E_{Lab}$=32.0 MeV. The parameters of the latter are shown in table \ref{param2nd}.}
\label{couplingpotcc}
\end{figure}

The limitations of the standard coupled-channels method, on the one hand, and
the {\it oblate} character of the $^{12}$C and the {\it prolate} character of
the compound nucleus $^{24}$Mg, on the other hand, have motivated us to use a
second-derivative coupling potential. In order to describe the above-mentioned
configuration, the coupling potential must be {\it oblate} (attractive) when
two $^{12}$C nuclei are at
large distances and must be {\it prolate} (repulsive) when they are at short
distances. The standard and the new coupling potential are shown in figure
\ref{couplingpotcc}.

One possible interpretation of such a second-derivative coupling potential
can be made if we express the total
potential as a function of the radii for different orientations of
the two colliding $^{12}$C nuclei. If $\theta_{P,T}$ are the angles between
the symmetry axes and the axis
joining the centers of the projectile and target, then the total potential,
as an approximation,
can be expressed in the following way:
\begin{eqnarray}
V(r)=V_{N}+\beta_{2}R\frac{dV_{C}}{dR}(Y_{20}(\theta_{P}, \phi_{P})+Y_{20}(\theta_{T}, \phi_{T}))+ \beta_{2}^{2}R^{2}\frac{d^{2}V_{C}}{dR^{2}}(Y_{20}(\theta_{P}, \phi_{P})+Y_{20}(\theta_{T}, \phi_{T}))
\label{orientfor}
\end{eqnarray}
where $V_{N}$ is the nuclear potential and $V_{C}$ is the new second-derivative
coupling potential. The final term is due to the mutual excitation.

\begin{figure}
\epsfxsize=9.0cm
\vskip-1.5cm
\centerline{\epsfbox{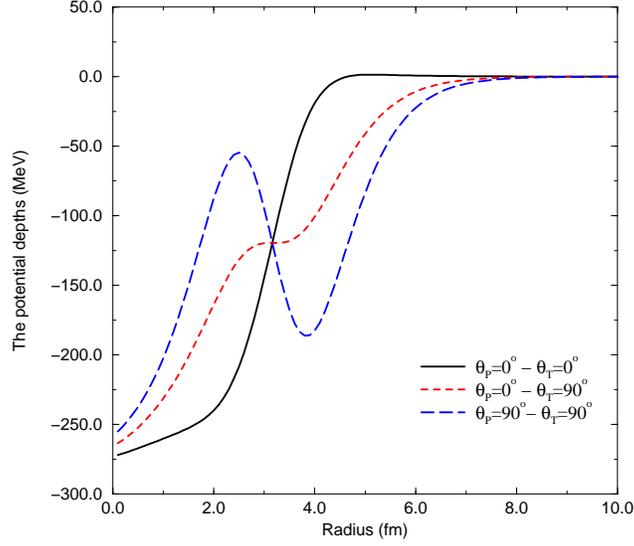}}
\vskip-3.5cm
\caption{The orientation potentials of two nuclei at different angles including the hexadecupole deformation of $^{12}$C.}
    \label{orient}
\end{figure}

The result for the  $^{12}$C+$^{12}$C system  is shown
in figure \ref{orient}. A second local minimum is observed in the
interaction potential for certain orientations. This feature, included only in an
{\it ad hoc} way in the work of Ordo\~{n}ez {\it et al}
\cite{Ord 86}, has not been taken into account in the standard coupled-channels
calculations. To investigate this minimum, we looked at the total inverted potential,
{\it i. e.} the dynamical polarization potential (DPP) plus the bare potential,
obtained by the inversion of the S-Matrix \cite{Boz3}.  Our analysis suggests that
the new coupling potential points to the presence of
the super-deformed configurations in the compound nucleus $^{24}$Mg,
as it has been speculated \cite{Rae88,Fes92}.

The real and imaginary potentials in these new calculations have
the same shapes as in previous calculations (see equation \ref{realpot}
and \ref{imagpot}) and the parameters of the potentials are displayed in
table \ref{param2nd}. We have analysed the experimental in the same energy range.

\section{Results}
\label{resnew}

The results of the analyses using the new coupling potential are displayed in
figures \ref{ground1} and \ref{ground2} for the ground
state, in figure \ref{single} for the first excited state
(single-2$^{+}$) and in figure \ref{mutual} for the mutual
excited state (mutual-2$^{+}$).

The agreement is very good for the
elastic scattering, single-2$^{+}$ and mutual-2$^{+}$ excitation inelastic scattering
data over the whole energy range studied. The theoretical predictions of the
magnitudes and the phase of the oscillations for the single-2$^{+}$ and
mutual-2$^{+}$ excitations inelastic scattering data, which have been the major
outstanding problems of this reaction, are in a very good agreement with the empirical values.
This new coupling potential has made a substantial improvement at
all the energies considered.

\begin{figure}
\epsfxsize 10.5cm \centerline{\epsfbox{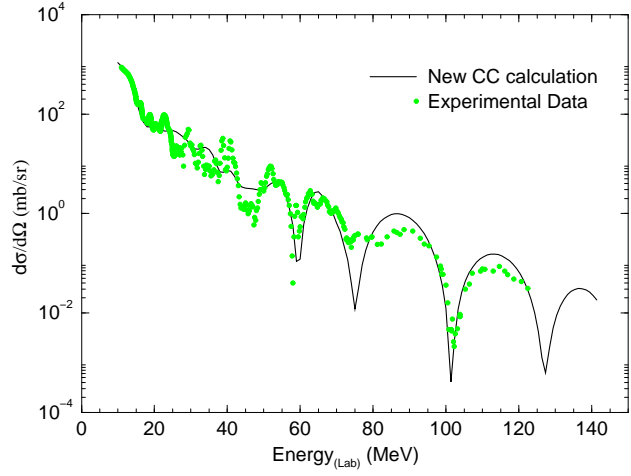}}
\vskip-8.0cm
    \caption{90$^{\circ}$ elastic scattering excitation function, obtained using new coupling potential with the empirical $\beta$ value ($\beta_{2}^{C}$=$\beta_{2}^{N}$=-0.6).}
\label{excfun}
\end{figure}

The 90$^{\circ}$ elastic scattering excitation function is also
analysed and the result is shown in figure \ref{excfun}.
The agreement with the experimental data is excellent
over the whole energy range.

Table \ref{param2nd} indicates that the parameters are almost constant
(1 to 3 $\%$ changes) away from the resonance regions. However, at
certain energies in the energy
range $E_{Lab}\sim$ 90 to 110.0~MeV, the parameters fluctuate.
We interpreted the fluctuations at small energies in table \ref{param2nd}
as the effect of the resonances observed by Cormier {\it et al}
\cite{Cor77,Cor78}, Chappell {\it et al} \cite{Cha95,Cha96,Cha98}
and Fulton {\it et al} \cite{Ful80}.
The changes of the potential parameters in the energy range $E_{Lab} \sim$
90 to 110.0 MeV might be related to resonances, which have not yet been
observed in the $^{12}$C+$^{12}$C system.  Within such an
interpretative scheme, one may infer that these resonances might be
associated with the single and mutual-4$^{+}$ excited states of $^{12}$C,
states which are strongly coupled to the ground state.

These predictions motivated us to run an experiment in this energy range.
The initial analyses of the experimental data indicate that the variation
of these parameters are not actually random since structures relating
to the 4$^{+}$ state of the $^{12}$C are seen in this energy range.
The detailed analyses and the full results will be given in the
forthcoming paper \cite{Bre99}.

This new, second-order coupling
potential, has also been applied successfully to the $^{16}$O+$^{28}$Si
and $^{12}$C+$^{24}$Mg systems \cite{Boz1,Boz2}.
This model has explained the experimental data successfully.
\section{Summary and Conclusion}
\label{summary}

We considered the elastic and inelastic scattering of the  $^{12}$C+$^{12}$C
system from 32.0 MeV to 126.7 MeV in the laboratory system. Although this reaction
has been one of the most extensively studied reaction over the last forty years,
there has been no global model that explains consistently the measured experimental
data over a wide energy range. In the introduction, we presented the problems that
this reaction manifests. We attempted to find a consistent solution to these problems.
However, within the standard coupled-channels method, we failed, as others did, to
describe certain aspects of the data, in particular, the single-2$^{+}$ and
mutual-2$^{+}$ excitation inelastic scattering data although the optical model
and coupled-channels models explain perfectly some aspects of the elastic scattering data.

As discussed in section \ref{res}, the standard coupled-channels method entails
that the coupling potential has the same energy-dependence as the central term.
However, our analysis reveals that the coupling potential has a vital importance
in explaining the experimental data for the reactions that involve at least one
strongly deformed nucleus and that there is no reason for the coupling potential
to have the same energy dependence as the central potential. This may explain the
failure of the standard coupled-channels calculations.

The comparison of the results obtained using the standard and new coupled-channels
calculations indicates that a global solution to
the problems of the scattering observables of this reaction
over a wide energy range (32.0 MeV to 126.7 MeV) with little energy-dependence on
the potentials has been provided
by this new coupling potential. The significance of the new approach
should be underlined because it does not only fit the present
experimental data, but it also leads to other novel and testable predictions.
To our knowledge, this has not been yet achieved over such a wide energy range. Studies
using this new coupling potential may also lead to new insights into the
formalism and a new interpretation of such reactions.
\section{Acknowledgments}
Authors wish to thank B. Buck, A. M. Merchant, Y. Nedjadi,
S. Ait-Tahar, R. Mackintosh, B. R. Fulton, G. R. Satchler and
D. M. Brink for valuable discussions and providing some data. I. Boztosun
also would like to thank the Turkish Council of Higher Education (Y\"{O}K) and Erciyes University, Turkey, for their financial support.

\tighten

\begin{table}[h]
\caption{The parameters of the potentials required to fit the
90$^{\circ}$ elastic excitation function, displayed in figure \ref{exc1}.}
\label{param}
\begin{center}
\begin{tabular}{llllll}
$V_{N}$ & $R_{N}$ & $a_{N}$ & $W$ & $R_{W}$ & $a_{W}$ \\
 (MeV) & (fm) & (fm) & (MeV) & (fm) & (fm)  \\     \hline
  345.0 & 3.62 & 1.60 & eq. (\ref{imag}) & 5.50 & 0.51 \\
\end{tabular}
\end{center}
\end{table}
\begin{table}
 \caption{The numerical values of the potentials used in the {\it standard coupled-channels} calculations. $V_{N}$, $r_{N}$ and $a_{N}$ stand for
the depth, radius and diffuseness of the real potential respectively
and W, $r_{W}$ and $a_{W}$ stand for the depth, radius and diffuseness of the imaginary potential respectively.}
\label{paramws2}
\begin{center}
\begin{tabular}{lllllllll}
$E_{Lab}$ & $V_{N}$ & $r_{N}$ & $a_{N}$ & $W$ & $r_{W}$ & $a_{W}$ \\
 (MeV) & (MeV) & (fm) & (fm) & (MeV) & (fm) & (fm) \\     \hline
     32.0 & 290.0 &   0.80 & 1.30 & 3.0 & 1.20 & 0.51 \\
     40.0 & 290.0 &   0.79 & 1.28 & 3.6 & 1.20 & 0.51 \\
     45.0 & 290.0 &   0.80 & 1.15 & 3.8 & 1.20 & 0.51 \\
     49.0 & 290.0 &   0.79 & 1.23 & 4.2 & 1.20 & 0.51 \\
     50.0 & 290.0 &   0.80 & 1.21 & 4.5 & 1.20 & 0.51 \\
     55.0 & 290.0 &   0.80 & 1.15 & 5.0 & 1.20 & 0.51 \\
     57.75 & 290.0 &   0.81 & 1.35 & 6.3 & 1.20 & 0.51 \\
     60.0 & 290.0 &   0.80 & 1.30 & 6.6 & 1.20 & 0.51 \\
     65.0 & 290.0 &   0.79 & 1.43 & 7.0 & 1.20 & 0.51 \\
     70.7 & 290.0 &   0.81 & 1.20 & 8.5 & 1.20 & 0.51 \\
     78.8 & 290.0 &   0.81 & 1.30 & 9.5 & 1.20 & 0.51 \\
     93.8 & 290.0 &   0.82 & 1.35 & 12.0 & 1.20 & 0.51 \\
     98.2 & 290.0 &   0.81 & 1.30 & 12.5 & 1.20 & 0.51 \\
    102.1 & 290.0 &   0.81 & 1.30 & 14.0 & 1.20 & 0.51 \\
    105.0 & 290.0 &   0.81 & 1.30 & 14.4 & 1.20 & 0.51 \\
     112.0 & 290.0 &   0.80 & 1.30 & 13.0 & 1.20 & 0.51 \\
     117.1 & 290.0 &   0.80 & 1.35 & 14.0 & 1.20 & 0.51 \\
     121.6 & 290.0 &   0.80 & 1.35 & 14.1 & 1.20 & 0.51 \\
     126.7 & 290.0 &   0.81 & 1.30 & 14.2 & 1.20 & 0.51 \\
\end{tabular}
\end{center}
\end{table}
\begin{table}
 \caption{The numerical values of the potentials used in the {\it new coupled-channels} calculations. $W$ denotes the imaginary potential. $V_{N}$, $r_{N}$ and $a_{N}$ stand for the depth, radius and diffuseness of the real potential respectively and $r_{C}$ and $a_{C}$ stand for the radius and diffuseness of the coupling potential respectively ($V_{C}$=210.0 MeV).}
\label{param2nd}
\begin{center}
\begin{tabular}{lllllllll}
$E_{Lab}$ & $V_{N}$ & $r_{N}$ & $a_{N}$ & $W$ & $r_{C}$ & $a_{C}$  \\
 (MeV) & (MeV) & (fm) & (fm) & (MeV) & (fm) & (fm)  \\     \hline
32.0 & 290.0 & 0.804 & 1.19 & 2.21 & 0.69 & 0.70 \\
40.0 & 288.0 & 0.806 & 1.28 & 2.40 & 0.69 & 0.70 \\
45.0 & 290.0 & 0.809 & 1.28 & 2.97 & 0.69 & 0.70 \\
49.0 & 290.0 & 0.810 & 1.28 & 3.07 & 0.69 & 0.70 \\
50.0 & 290.0 & 0.813 & 1.24 & 3.07 & 0.69 & 0.70 \\
55.0 & 290.0 & 0.813 & 1.26 & 3.17 & 0.69 & 0.70 \\
57.75 & 290.0 & 0.813 & 1.26 & 3.17 & 0.69 & 0.70 \\
60.0 & 290.0 & 0.813 & 1.28 & 3.37 & 0.69 & 0.70 \\
65.0 & 290.0 & 0.811 & 1.28 & 3.57 & 0.69 & 0.70 \\
70.7 & 289.0 & 0.799 & 1.29 & 3.71 & 0.69 & 0.70 \\
78.8 & 287.0 & 0.785 & 1.28 & 5.50 & 0.68 & 0.70 \\
93.8 & 292.0 & 0.790 & 1.34 & 11.9 & 0.67 & 0.67 \\
98.2 & 289.0 & 0.785 & 1.27 & 11.5 & 0.66 & 0.65 \\
102.1 & 289.0 & 0.810 & 1.33 & 11.5 & 0.65 & 0.63 \\
105.0 & 289.0 & 0.810 & 1.37 & 11.5 & 0.66 & 0.66 \\
112.0 & 287.0 & 0.800 & 1.28 & 13.8 & 0.68 & 0.67 \\
117.1 & 290.0 & 0.810 & 1.32 & 14.7 & 0.69 & 0.68 \\
121.6 & 290.0 & 0.810 & 1.33 & 15.3 & 0.68 & 0.67 \\
126.7 & 288.0 & 0.795 & 1.30 & 17.3 & 0.66 & 0.67 \\
\end{tabular}
\end{center}
\end{table}


\end{document}